# Environmental Impacts of Personal Protective Clothing Used to Combat COVID-19

***Mohammad Abbas Uddin[1], Shaila Afroj[2], Tahmid Hasan[3],***

***Chris Carr[4], Kostya S Novoselov[5,6,7], and Nazmul Karim[2]****

[1]*Department of Dyes and Chemicals Engineering, Bangladesh University of Textiles, Tejgaon, Dhaka-1208, Bangladesh*

[2]*Centre for Print Research (CFPR), The University of West of England Bristol, Frenchay Campus, Bristol BS16 1QY, UK*

[3]*Department of Environmental Science and Engineering, Bangladesh University of Textiles, Tejgaon, Dhaka-1208, Bangladesh.*

[4]*Clothworkers' Centre for Textile Materials Innovation for Healthcare, School of Design, University of Leeds, Leeds LS2 9JT, UK*

[5]*Department of Materials Science and Engineering, National University of Singapore, Singapore.*

[6] *Institute for Functional Intelligent Materials, National University of Singapore, Singapore.*

[7]*Chongqing 2D Materials Institute, Liangjiang New Area, Chongqing, 400714 China*

***Email:** nazmul.karim@uwe.ac.uk*

*Personal protective clothing is critical to shield users from highly infectious diseases including COVID-19. Such clothing is predominantly single-use, made of plastic-based synthetic fibres such as polypropylene and polyester, low cost and able to provide protection against pathogens. However, the environmental impacts of synthetic fibre-based clothing are significant and well-documented. Despite growing environmental concerns with single-use plastic-based protective clothing, the recent COVID-19 pandemic has seen a significant increase in their use, that could result in a further surge of oceanic plastic pollution, adding to mass of plastic waste that already threatens marine life. In this review, we briefly discuss the nature of the raw materials involved in the production of such clothing, as well as manufacturing techniques and the PPE supply chain. We identify the environmental impacts at critical points in the protective clothing value chain from production to consumption, focusing on water use, chemical pollution, $CO_2$ emissions and waste. On the basis of these environmental impacts, we outline the need for fundamental changes in the business model, including increased usage of reusable protective clothing, addressing supply chain "bottlenecks", establishing better waste management, and the use of sustainable materials and processes without associated environmental problems.*

## 1. Introduction

The worldwide demand for personal protective equipment (PPE) has increased in recent months to an unprecedented level, due to the COVID-19 pandemic.[1] As a result, the manufacturing and distribution of single-use PPE has seen a huge growth, notably in surgical masks and gowns which are made from plastic-based polypropylene nonwoven fabrics.[2] The World Health Organisation (WHO) has prescribed a variety of measures to contain and prevent the spread of viruses to the community and health care workers,[3] which includes community lockdown, travel restrictions, social distancing, isolation, hand sanitising and the mass wearing of disposable face masks and gloves.[4] Within this strategy, the use of PPE is the critical component to protect healthcare workers (HCWs), patients, front-line workers and the mass population from highly infectious diseases such as COVID-19.[5] Furthermore, the European Centre for Disease Prevention and Control (ECDC) estimated that health services would require 14 to 24 separate sets of PPE every day for each confirmed COVID-19 case, depending on the severity of the symptoms.[6] In March 2020, the modelling carried out by WHO indicated that there would be a global need for ~89 million medical masks, ~76 million examination gloves, and ~1.6 million pairs of goggles in each month, in response to the pandemic, Figure 1a.[7] However, most PPE items, such as masks and gloves, are made of plastics and single-use, meaning they will need to be disposed after each use, leading to the creation of large volumes of waste.[8] Additionally, the daily consumption of single-use face-masks by the general population will increase non-recyclable plastic waste, and have a detrimental impact on the environment, as currently there is no infrastructure in place for the safe and environmentally friendly disposal of potentially contaminated single-use face masks used by the general population.[8, 9]

PPE is defined as "*equipment worn to minimise exposure to hazards that cause serious workplace injuries and illnesses. These injuries and illnesses may result from contact with chemical, radiological, physical, electrical, mechanical, or other workplace hazards*."[10] Amongst the PPEs, protective clothing is designed to protect the eye, face, head, leg, hand and arm, body, and hearing organs,[11] and is classified as Level A, B, C and D for the general population, where Level A offers the highest level of the skin, eye, and respiratory protection.[12] Personal protective clothing for medical or healthcare applications are used to mitigate the risks from exposure to hazardous substances, including body fluids and to minimise the risk of cross-infections.[13] Such single use protective clothing are made of synthetic fibres such as polypropylene and polyester, due to their low-cost, hydrophobic nature and better barrier properties.[13, 14] However, the production of synthetic fibres from fossil oil is associated with significant carbon emission. For example, synthetic fibres are responsible for two-thirds of the total ~10% global carbon emissions associated with textile materials.[13] Additionally, such fabrics are not readily biodegradable, remaining in the environment (air, soil or sea) for hundreds of years and can have significant impact on human health, Figure 1b-e .[13]

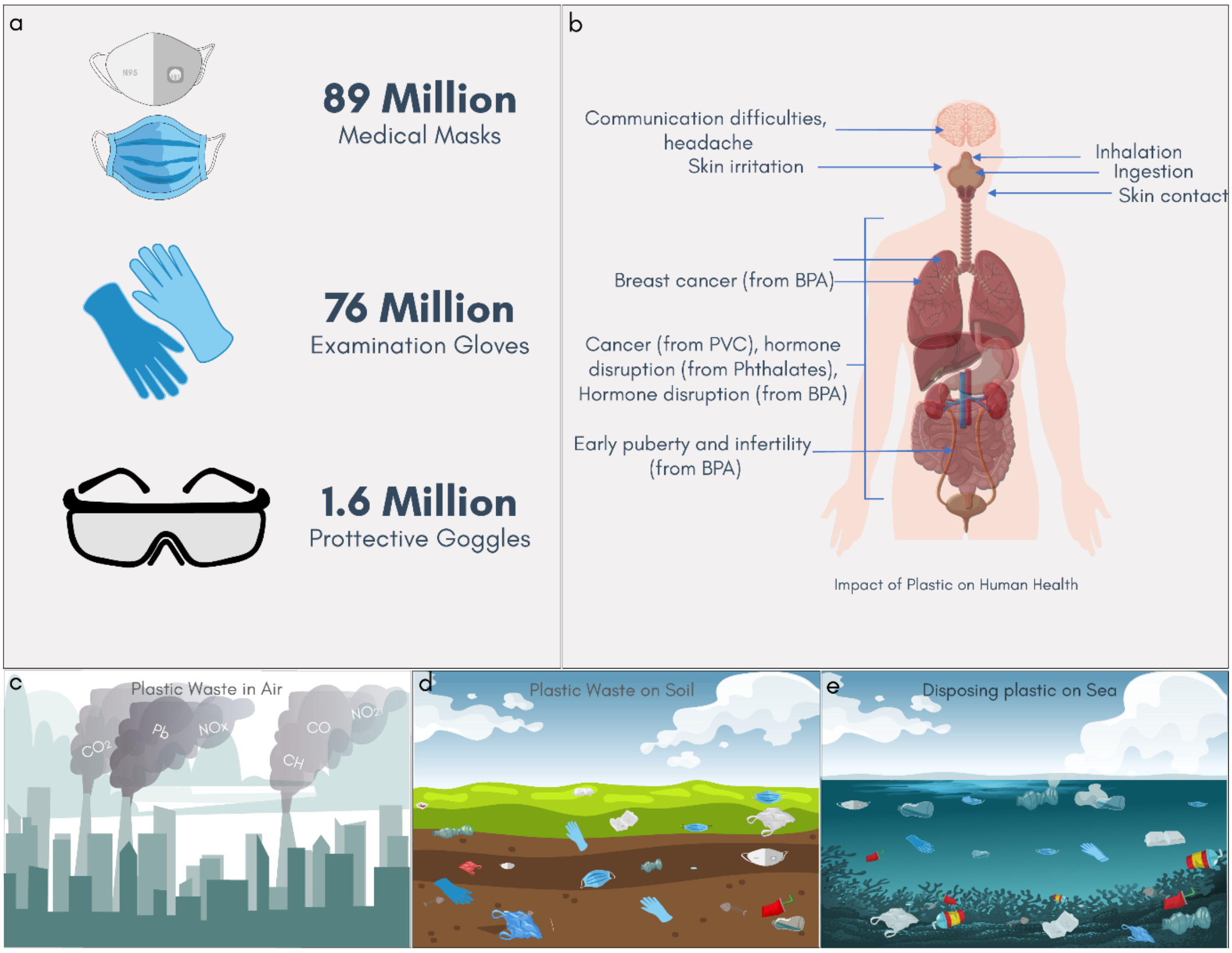

***Figure 1. PPE and Health: The hidden cost of plastic-based PPE waste. a)*** *Estimated Number of PPE (medical masks, examination gloves and protective goggles) needed per month during COVID-19 pandemic according to a model carried out by WHO in March 2020.* ***b)*** *The impact of plastic on human health. Plastic-based PPE in* ***c)*** *air,* ***d)*** *soil and* ***e)*** *sea.*

The recent surge in single-use protective clothing consumption due to COVID-19 represents the key environmental threat. Indeed, considerations of pollution and waste were not of primary concern for manufacturers and consumers, with the primary focus being on protection from the highly infectious COVID-19 pathogens. However, with growing warnings from the environmentalists and increased public awareness of the climate crisis and sustainability in general, the industry (manufacturers, suppliers and consumers alike) will be forced to seek more sustainable and "circular" protective clothing and consider their environmental impacts. In this review, we provide a brief overview about raw materials for personal protective clothing and their manufacturing processes. We then outline PPE global supply chains, and pre-COVID-19 and during COVID-19 market size. We discuss the environmental impacts of single-use personal protective clothing, specifically, the global map for single-use-plastic waste, pollutions (aquatic, marine and chemical), and its environmental footprints before and during COVID-19. Finally, we present our recommendations and perspectives on how the products or technology can be changed to become more sustainable, including decreasing the use of single-use protective clothing and their waste, and moving towards smart, sustainable and reusable protective garment usage and embedding a longer lifetime framework.

## 2. Protective Clothing: Raw Materials and Manufacturing Processes

To minimise the exposure to infectious microorganisms or hazardous materials in medical environments, several different types of medical clothing products are used, including coveralls, footwear covers, full-body suits, independent sleeves, scrubs, surgical gowns, surgical masks, and scrub hats.[15] Single-use nonwoven fabrics are popular choice for such clothing, as they provide excellent protection against fluids (blood and other bodily fluids) and pathogens, as well as maintaining garment breathability and comfort.[16] Petrochemical-based synthetic fibres (such as polypropylene, polyester, and polyethylene) are typically used for single-use protective clothing, which have been engineered to achieve the desired performance properties based on fibre type, bonding processes, and fabric finishes (chemical or physical).

### 2.1 Raw materials of plastics

The raw material for any protective clothing is fibre whether from natural or synthetic sources. Following the recognition of macromolecules by W. H. Carothers in 1928, and the subsequent development of the first synthetic fibre, polyamide 66 in 1935, and its commercial introduction as nylon in 1938, the growth of the use of synthetic fibre has been exponential.[17] Synthetic fibres are essentially polymeric materials, and depending on their use, could be generically classified as 'plastics', the quintessential product material for our modern lifestyle. Due to the ready availability of raw materials (derived from the petrochemical industry), tailor-made physio-chemical properties (e.g. flexibility, lightweight, durability), and production in economic scale, plastics quickly started to dominate many industrial sectors such as healthcare, packaging, agriculture, and fisheries, surpassing any other manmade materials.[18] Other than fossil fuel sources, plastic materials can also be produced from renewable sources (e.g. sugar cane, starch, vegetable oils) or mineral base (salt).[19] According to the Plastics Europe market research group,[20] total worldwide plastic production was ~368 million metric tons in 2019, (with slight reduction of approximately 0.3% in 2020) and Europe consumed ~50.7 million tons of the total plastic production. Asia is the leading consumer of plastics with ~51% of total global consumption (China ~30%, Japan ~4% and rest of Asia ~17%), followed by Europe (~17%), NAFTA (~18%), Middle East and Africa (~7%), Latin America (~4%) and Eastern Europe (~3%), **Figure 2a.** The most common polymers, which account for about ~82% of European plastic demand in 2019, are polyethylene (PE), polypropylene (PP), poly(vinyl chloride) (PVC), polystyrene (PS), poly(ethylene terephthalate) (PET) and polyurethane (PU), Figure 2b.[20] Table 1 shows that the most commonly used synthetic fibres for protective clothing applications are: polypropylene (PP), low-density and linear low- density polyethylene (LDPE and LLDPE), and polyester (PET). The properties of such fibres (e.g. inherent absorbency) determine the level of protection against the contaminants/microorganism, with microfibres usually preferred when a higher level of protection needed.

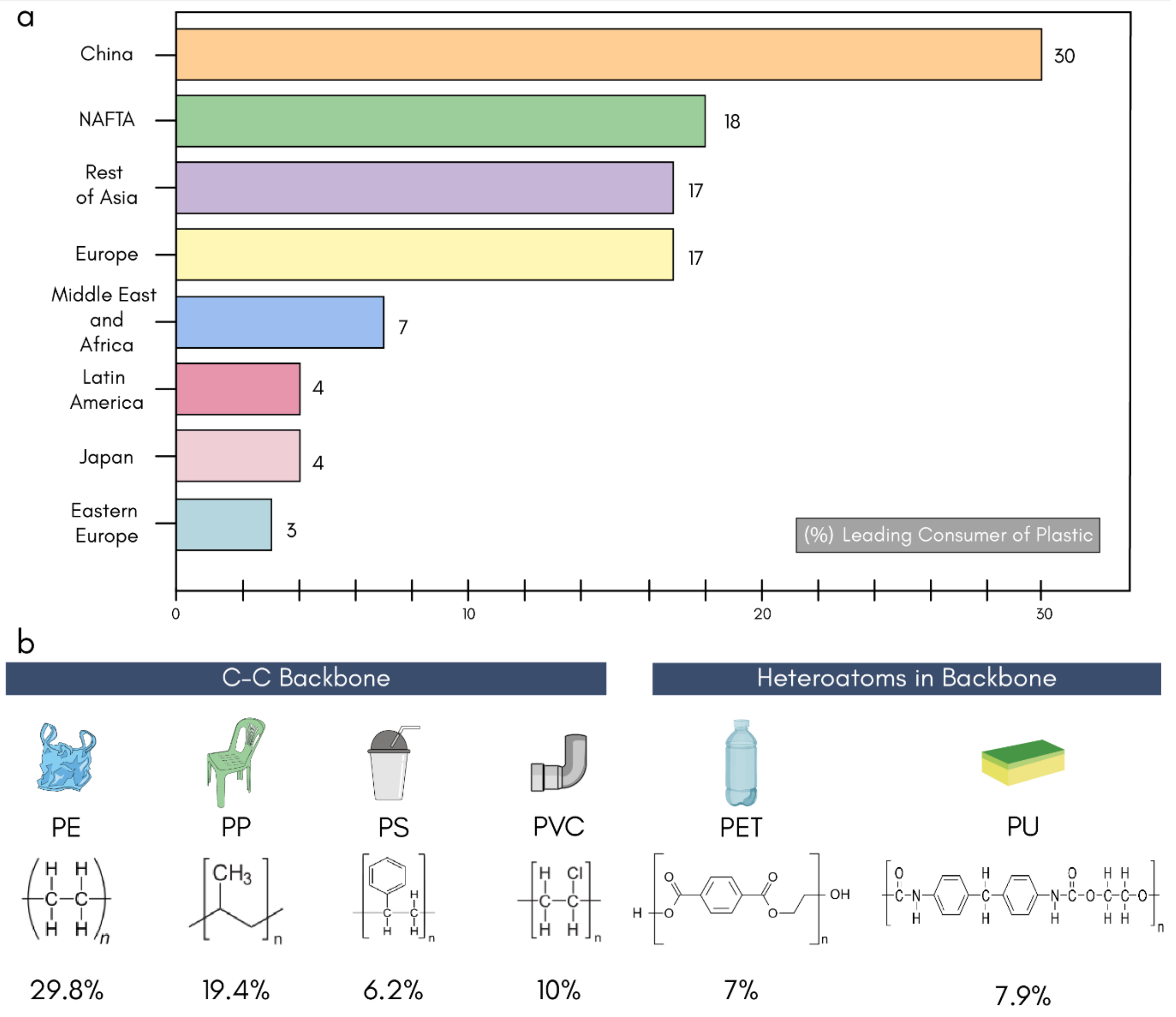

***Figure 2 Plastic consumers and polymers. a)*** *Leading plastic consuming countries and continents in the world.* ***b)*** *Global demand for polymer materials and specific contributions of PE – polyethylene, PP – polypropylene, PS – polystyrene, PVC – poly(vinyl chloride), PET – poly(ethylene terephthalate), PU – polyurethane) within the total EU demand for plastic of 50.7 million tons.*[20]

*Table 1 Single-use PPEs: polymers, manufacturing processes and properties*

| PPEs | Polymers | Manufacturing process | Key properties | Quality control | Ref |
|---|---|---|---|---|---|
| Surgical mask | Polypropylene, polyurethane, polyacrylonitrile, polystyrene, polycarbonate and LDPE or polyester | Spunbond-meltblown-spunbond | Vapour and liquid absorption and tensile strength | EN 14683 Type IIR performance<br><br>ASTM F2100 level 2 or level 3 | [21] |
| FFP2/N95 Mask | LDPE and modacrylic | Spunbond-meltblown-spunbond | Protection against airborne and flow rate | NIOSH N95, EN149 FFP2, | [2] |
| Nitrile gloves (nitrile butadiene rubber) | Acrylonitrile and other copolymers | Polymerisation | Chemical resistance and tensile strength | EU standard directive 93/42/EEC Class I, EN 455<br>EU standard directive 89/686/EEC Category III, EN 374<br>ANSI/ISEA 105-201<br>ASTM D6319-10 | [22] |
| Single-use Apron/ gown | Polypropylene and polyester | Spunbond-meltblown-spunbond | Absorbency, barrier and chemical resistance | EN 13795 high performance level, or AAMI level 3 performance, or equivalent;<br>or<br>AAMI PB70 level 4 performance, or equivalent | [13, 22, 23] |
| Surgical drapes | Polypropylene | Spunbond | Barrier to liquid, microorganism, humidity | EN 13795 for fabric, ISO 16603 class 3 | [23, 24] |
| | | spunbond-meltblown-spunbond | | exposure pressure, or equivalent; or Option 2: ISO 16604 class 2 exposure pressure, or equivalent | |
| Face shield | Polycarbonate, propionate, acetate, polyvinyl chloride, and polyethylene terephthalate glycol | Extrusion and injection moulding | Impact resistance, optical quality and chemical resistance | EU standard directive 86/686/EEC, EN 166/2002<br><br>ANSI/ISEA Z87.1-2010 | 25 |
| Shoe and head cover | Polypropylene and polyethylene | | Durability and anti-skid | | 26 |
| Goggles and safety covers | Cellulose acetate, cellulose propionate and polycarbonate | Injection moulding and surface treatment | Particle resistance and impact resistance | EU standard directive 86/686/EEC, EN 166/2002<br><br>ANSI/ISEA Z87.1-2010 | 27 |

### 2.2 Fabric manufacturing

Single-use protective clothing is predominantly nonwoven in construction, as non-woven fabric facilitates relatively fast and cheap manufacturing, high levels of sterility, and infection control. Such nonwoven fabrics are typically made from polypropylene, and usually have a spunbond–meltblown–spunbond (SMS) construction in the final products. Nonwoven fabrics are formed as a web by directly entangling textile fibres together, which works as a base for further bonding to increase the fabric's strength. Surface modification can be performed through mechanical treatment or coating, Figure 3.[28] A detailed description of fabric manufacturing (both woven, knit and nonwoven) and anti-microbial finishing techniques can be found in our previous review.[2]

Most commonly used web formation technologies for manufacturing nonwoven fabrics are: dry-laid, wet-laid and spun-laid. In dry-laid technology, carding or air laying of the fibres are used to produce nonwoven web. In contrast, the wet-laid technology uses a similar technique as papermaking to manufacture nonwoven fabric from a slurry of fibres and water.[29, 30] However, wet-laid nonwovens are differentiated from wet-laid papers by having more than 30% of its fibres with a length to diameter ratio greater than 300, and a density less than 0.40 g/cm$^3$.[4, 31] Nonwoven webs can be formed from natural and manmade fibres in staple form using these two techniques.[30] The other web formation technique is the spun-laid process, which uses melt spinning technique to form the web, thus eliminating the expensive transformation of polymers into staple fibres. In the spun-laid process, only the synthetic fibres, predominantly high and broad molecular weight thermoplastic polymers such as polypropylene, polyester, and polyamide, are extruded through spinneret as endless filaments, which are then cooled and stretched by air, and are laid down in a continuous process. Several methods can be used to produce spun-laid nonwoven fabrics including spun-bond, melt-blown, aperture films, and the many-layered combinations.[32, 33] Among them, the melt-blown process (Figure 3a) provides the advantages of better filament distribution, better filtration *via* smaller pores between the fibres, softer feel, and also the possibility of manufacturing lighter weight fabrics. The difference between spun-laid and melt-blown processes is that the latter have a higher melt flow index of the polymer with lower throughput, which results in the manufacture of very fine fibres.[28, 31, 34]

The strength of the nonwoven web is increased by consolidating the fibres using a thermal, mechanical or chemical bonding processes. The most common web bonding for producing medical textiles is thermal bonding (Figure 3b),[29] which is achieved *via* melting thermoplastic fibres or their blends (often containing *binder fibres*). The binder fibre component (usually ~5–50 wt.-%) can be in powder, film, low melt webs, and hot melts form for disposable and durable products.[35] For thermal bonding, the webs are either moved in between heated calendar rollers or hot air is blown through the web.

Mechanical bonding is the oldest web bonding process produced through needle punching, hydroentanglement or stitching. Needles or high-pressure water jet are passed through the web to increase the physical entanglement of the fibres. Such hydroentangled fabrics have been used for surgical gowns, scrub suits, sheet and drapes due to their excellent comfort and softness, however they have low barrier properties.[36] The chemical web bonding takes place *via* liquid-based chemical, which works as a binder. The chemical bonding is a popular method, due to the availability of extensive range adhesive, the product durability and a broad range of properties that can be engineered in the fabrics. The bonding agent can be applied *via* saturation bonding, foam bonding, print bonding, coating or scraper bonding, and solution and partial solution bonding.[37]

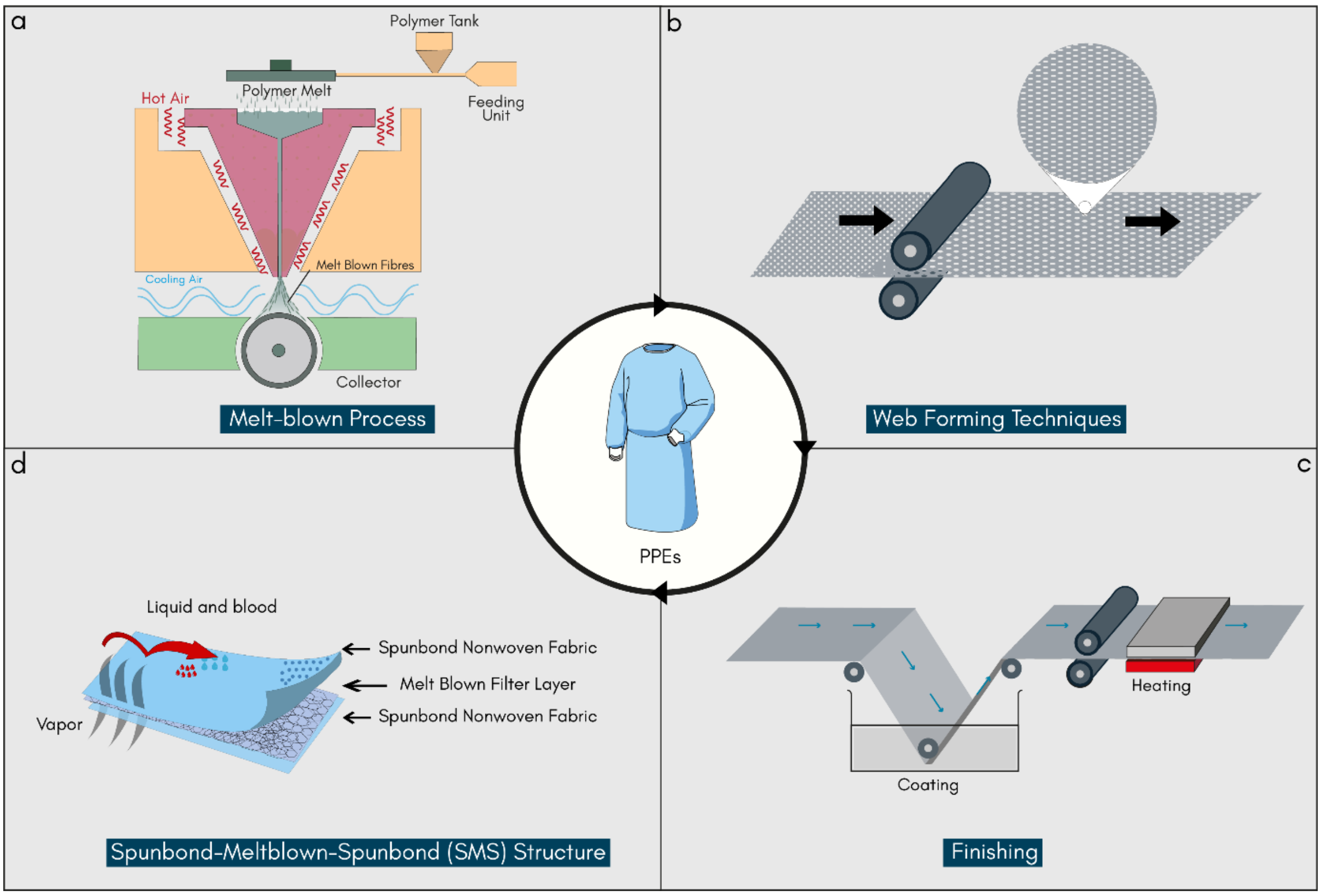

***Figure 3 Protective clothing manufacturing.*** *a) Melt-blown process. b) Thermal bonding technique for web formation. c) Pad-dry-cure finishing technique to impart antimicrobial or other functional finishes and d) Three-layer spunbond-meltblown-spunbond (SMS) structure which is mostly commonly used for personal protective clothing to protect against highly infectious diseases. (a, c, and d are reproduced with permission[2] and further permissions related to the material excerpted should be directed to the ACS)*

The finishing of nonwoven fabrics occurs as the last stage, mainly following traditional textile finishing techniques: dry finish and wet finish (Figure 3c). However, there are many nonwoven fabrics which do not undergo any finishing at all before packaging. Wet finishing includes colouration, washing, coating, and printing, while dry finishing includes calendering, embossing and emerising. The choice of finishing processes depends on the specific end-use application. In the hygiene and medical industry, nonwoven fabrics are often impregnated with detergents, cleaning agents, finishing agents or other lotions.[28, 38]

### *2.3 Protective garment*

Compared to traditional garment making, PPE manufacturing requires fewer stages but may rely on some specialised machinery. Ultrasonic welding and sewing machines are required to stitch at the edge for masks and gowns. In many cases, several layers of nonwoven fabrics are used to provide different functionalities as required by the end users.[39] Additionally, different types of finishing could be applied in different layers depending on the end use requirements, such as SMS fabric for maximum breathability and high fluid repellency, Figure 3d. Surgical masks for healthcare applications require high bacterial filtration efficiency for maximum protection, therefore fibres, fabrics, and finishing are chosen according to the fibre's intrinsic properties and construction of the materials. For example, Type IIR masks have a slash-resistant finish in some layers of SMS configuration.[39] N95 respirators have extra filtration layers and are designed to have close facial fit, which assists in very efficient filtration of airborne particles.[2] Based on such special characteristics, which are incorporated for extra efficacy through an additional layer of finish, a product could be classified as PPE or medical device. For example, a glove could be of surgical use in a hospital or for laboratory use in a university.[40] Many standard gowns are made in layered spunbond-meltblown-spunbond (SMS) fabrics, which are available in different thicknesses to provide various level of protection.

## 3. Global Protective Clothing Supply Chains

Even before the COVID- 19 pandemic, the use of protective clothing was increasing due to increasing regulation in the workplace, greater industrial awareness of employee protection, and high economic growth in countries such as Japan, India, China, Germany, and the US. The global market for PPEs in 2019 was worth over $52.7 billion, which was expected to grow at a Compound Annual Growth Rate (CAGR) of 8.5% to over or over $92.5 billion by 2025, Figure 4.[41] Since the demand of the protective clothing is growing around the world, so is the demand/supply of associated textile fibre, and as a result, the relationship within the stakeholders of the textile supply chain has much more profound effect in the protective clothing market.

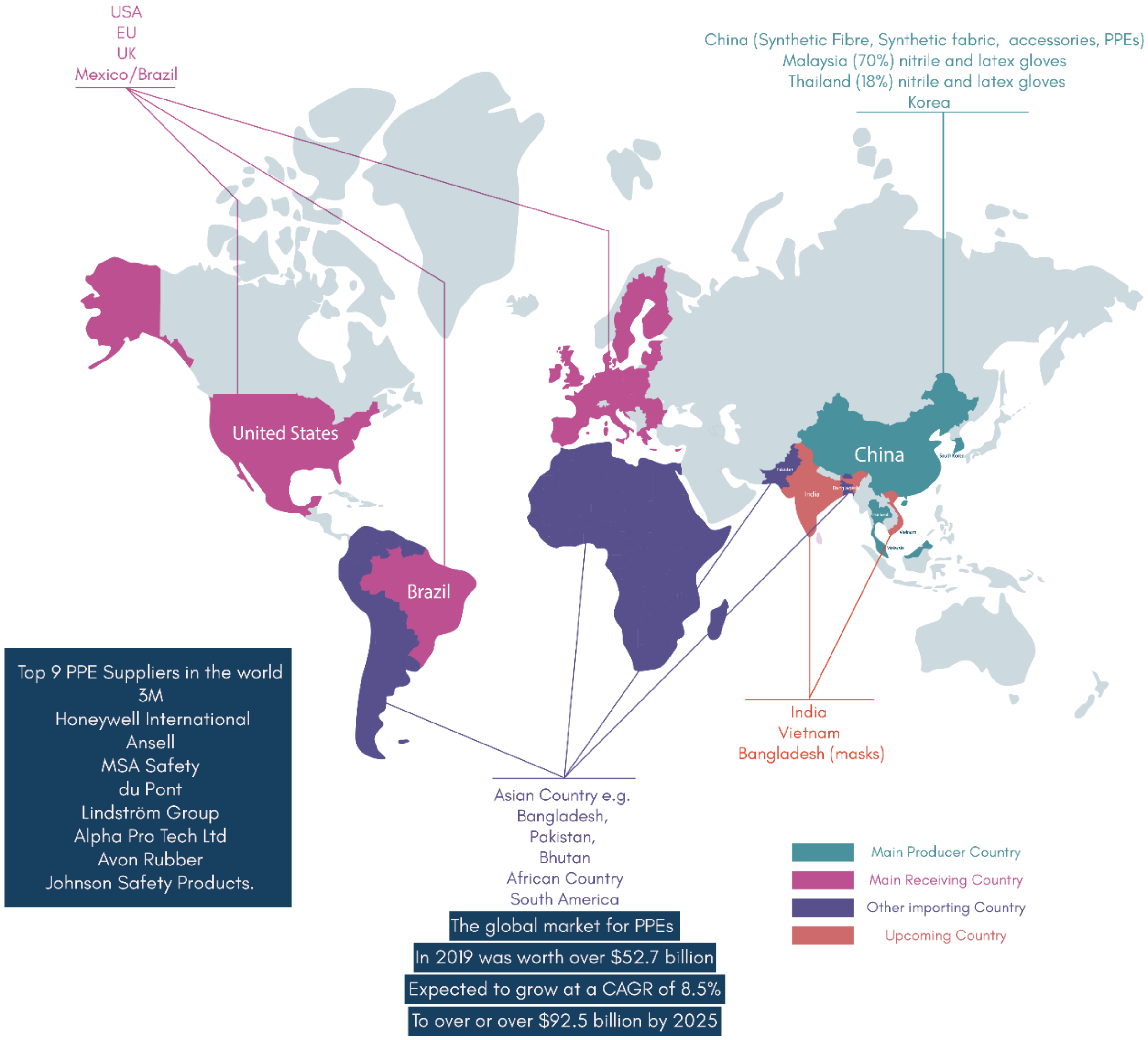

***Figure 4 Global protective clothing supply chain**. China is the main protective clothing manufacturing country. UK, USA, EU, Mexico and Brazil are receiving countries (leading consumer countries). Bangladesh, India and Bhutan are emerging manufacturing countries. The other countries such as Bangladesh, Pakistan, Bhutan, African countries and South American countries are importing countries of protective medical clothing.* [41, 42]

In general, any textile supply chain is characterised by a vertical set of globally dispersed industries: agriculture and farming for natural fibre production, petrochemical for synthetic fibre production, along with spinning, weaving/knitting, dyeing/finishing and apparel manufacture, and then logistics and distribution.[43, 44] Such complexity has further been increased in the case of protective clothing manufacturing, where local distributors with regular weekly supplies usually dominate the PPE supply chain. These distributors will either provide contracts directly to manufacturers or through a third party to manufacture PPE products.[45] Again, the distribution channels could also be divided based on direct/institutional sales or retail sales, where clients can buy PPEs directly from these distributors. Although the global protective clothing market has an extensive network of small and medium enterprises, the market is still dominated by leading brands. The largest PPE manufacturers in the world are 3M, Honeywell International, Ansell[46] along with MSA Safety, DuPont, Lindström Group, Alpha Pro Tech Ltd, Avon Rubber, and Johnson Safety Products.[42, 45] However, there is no primary data available on domestic production of PPEs by those companies.

The PPE supply chain is characterised by high geographic and regional concentration with three emerging regional clusters: Asia, Europe, and the US.[47] More than 70% of respiratory products used in the USA are manufactured in Asian countries such as China, Malaysia, Thailand and Korea. In addition, polymer raw materials, melt-blown fibres and accessories (e.g. nose clips) required to make N95 masks are mostly produced in China.[48] Thus, China is the manufacturing hub of most types of protective clothing along with the raw materials to produce them, such as synthetic fibres, fabrics, and accessories. This extensive influence throughout the supply chain also dominates the shipment and distribution channels. Other countries might be leading producers of other products, such as for single-use medical nitrile and latex gloves: Malaysia (70%), Thailand (18%) and China (10%).[49-51]
Up until now (26 August 2021), over 214 million COVID-19 patients and over 4.4 millon deaths in more than 222 countries were found and the number growing daily.[52] It is reported that as a consequence of the COVID-19 pandemic, the global production of healthcare PPE increased by at least 300% between 2019 and 2020,[53, 54] mainly driven by demand for masks. Before the COVID-19, the PPE market was dominated by distributors (~60% of PPE transactions in the US and 70% in the EU), which has been changed considerably during the COVID-19 where the government became the major PPE buyers in the US, EU, UK and China (increaed to ~60% government purchase from ~5% pre-COVID-19). These countries also increased their production drastically. For example, Europe has increased PPE (such as mask) production by 20 times. More than 3000 new PPE manufacturing industries form China entered into the market with 4000 existing manufacturers, which resulted in increased local production by ~1,000% for masks and 300-500% for gloves during the last quarter of 2020. China produces 200 million face masks a day, which is ten times higher than the monthly average in February 2020.[55]

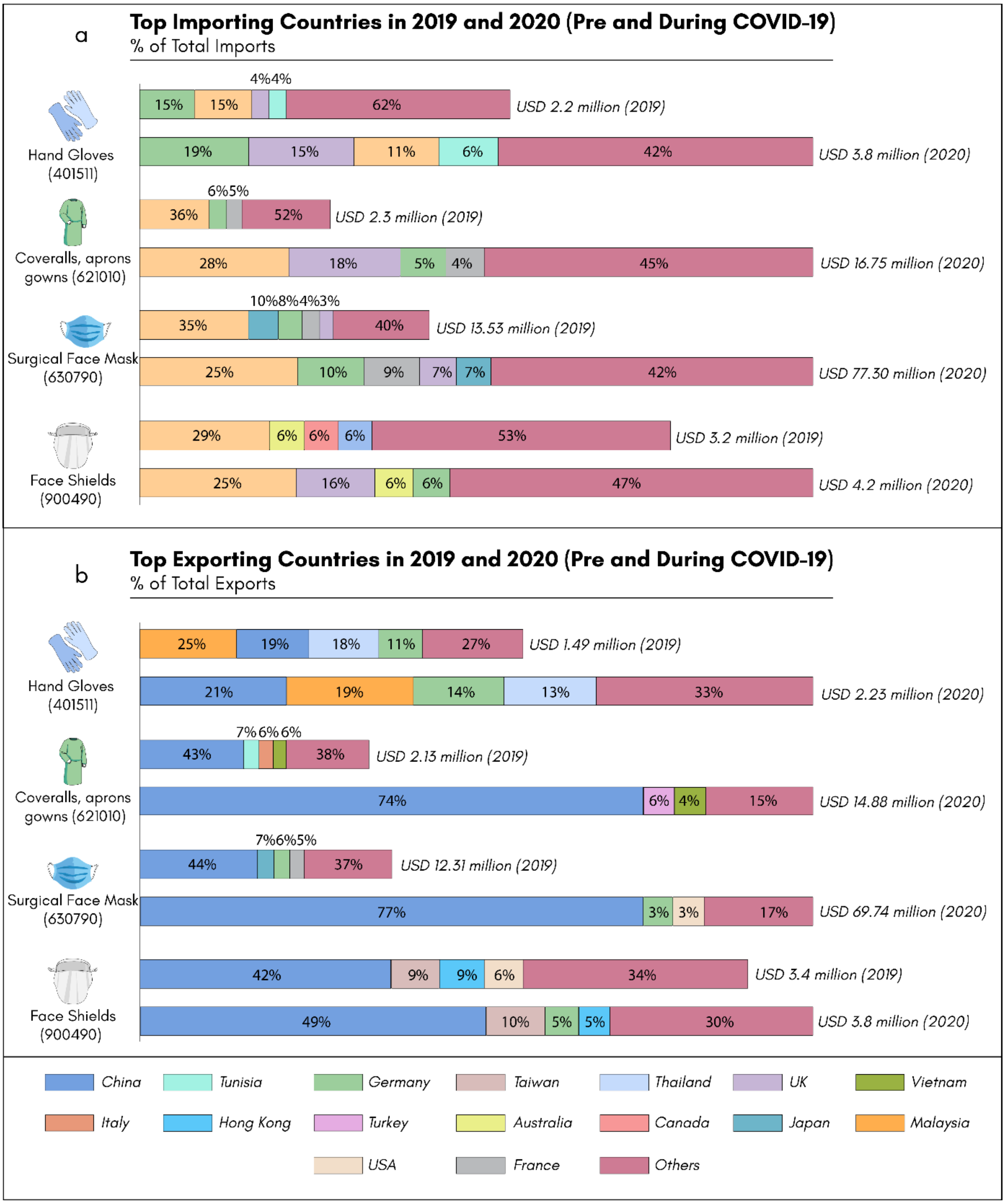

***Figure 5 Pre-COVID-19 and During COVID-19:*** *The global import and export market for four types of PPEs in 2019 (Pre-COVID-19) and 2020 (During COVID-19). a) Top importing countries b) top exporting countries. The USD value represent the total export/import in that particular year based on the six-digit HS Code (underneath each items). However, these six-digit code also includes other products based on the category given above.*[53, 54]

The traditional textile manufacturing countries like Bangladesh, India, Sri Lanka and Vietnam had a very limited PPE products before COVID-19. However they have increased their PPE production signifcantly since COVID-19 outbreak by modifying existing production line. The global demand of these PPEs is expected to continue in 2021 due to the unpredictability (e.g., new COVID-19 variants) of the crisis but is also expected to decline by ~50% in 2022 from the demand in 2021.[53] Based on the six-digit HS codes, Figure 5a,b show comparative analysis of import and export data for PPE products in 2019 (pre-COVID-19) and 2020 (during COVID-19).[53, 54]

Thus the demand for the PPEs became manyfolds in 2020 and it will not subside significantly until and unless pandemic can be contained.[56] To meet the demand China produces ~240 tons of medical waste daily in Wuhan alone, and Hong Kong residents use ~7 million masks single-use masks daily.[57] As prescribed by the WHO, almost all countries recommend using masks in enclosed spaces.[58] From February to August 2020, nearly 1.8 billion gloves were supplied to the UK National Health Service (NHS).[59] The UK government has allocated an extraordinary £15 bn for procuring masks, gowns and gloves to mitigate against COVID-19 in July 2020.[60] If everyone in the UK uses a single disposable mask every day it would create up to 66,000 tons waste in a year.[61] Thus, this heavy dependence on a few countries, and globally diverse supply chains has an unprecedented consequence, especially for high-value, high-risk products such as respirators and N95 masks.[46] Any rapid or unexpected surge in demand for PPEs, such as in the event of a public health emergency, puts a strain on the supply chain. This has been the case during the 2009 H1N1 influenza pandemic, the 2014 Ebola virus epidemic,[62] and also current COVID-19 pandemic.[2, 63] Various initiatives have been taken to curb this dependency. For example, the International Finance Corporation (IFC) has developed a PPE calculator to assess investments and working capital needs to switch from mainstream textiles to produce PPEs.[64] Some commercial technology-based companies are also assisting in retooling to PPE production through Industry 4.0 (I4.0) technology, and also providing *'PPE Manufacturing Matchmaking Program'* to connect their global network of manufacturers and suppliers.[65]

**4. Plastic Pollution linked to COVID-19 PPEs**

The environmental impacts due to plastic and plastic particles are well documented in the literature.[21, 66-71] However, this environmental impact has increased significantly with increasing production and consumption of single-use PPEs,[6] and the new emergence of mandatory face masks has not reduced the challenge of PPE pollution in the environment, be it Africa, Asia, EU, the US or elsewhere.[6, 72]

### *4.1 Global problem, Local impact*

Plastic contributes to climate change through greenhouse gas (GHG) emission, marine pollution, food security and freshwater scarcity.[72] To reduce the environmental impact of plastics, and plastic leakage, several initiatives and directives have been developed at international, national, and regional levels, including environmental taxes or bans on certain single-use plastics.[73] However, while the emergence of COVID-19 has caused some significant environmental improvements, for example, improved outdoor air quality and decreased number of smokers,[74] nevertheless, the pandemic has forced rapid and wide use of single-use plastic-based protective clothing by the mass population, and resulted in the accumulation of potentially infectious domestic solid waste streams.[1, 14]

The shift of single-use PPEs is mostly driven by potential cross-contamination and hygiene concerns.[73] Accordingly human health has been prioritised over environmental health, reduction policies and waste management strategies of plastics have recently been reversed or temporarily postponed.[14] Many governments have delayed restrictions of single-use PPEs such as Newfoundland and Labrador in Canada, New York and Oregon in the US, Portugal, England and Australia.[75] Even Senegal which bans single-use plastics including imports of plastic waste, acknowledges that enforcement of such measures during the COVID-19 pandemic is unlikely.[76] Similarly, California, New York, Maine, Massachusetts (USA) banned single-use plastic-based shopping bags some years ago, however it has again effectively reverted back to single-use bags to protect from COVID-19 infection.[77-79] Additionally, a dramatic fall in petroleum prices favoured the manufacturing of virgin plastics compared to the recycled plastics.[58] Thus, the environmental burden for the society has increased significantly in recent months.

### *4.2 Marine disposal of SUPs*

The presence of microplastics is ubiquitous in the marine environment worldwide.[80] Single-use plastics (SUPs) contribute to ~60-95% of global marine plastic pollution,[73] with ~50% of plastics in the ocean more than 30 years old. In 2015, it was found that ~90% of the plastic was over two years old.[81] Perhaps not surprisingly, the world's ocean floor is littered with an estimated ~14 million tonnes of microplastics.[82] PPEs are lightweight and can easily be carried out by wind or surface currents and quickly spread in the natural environment.[44] Plastic waste can be broken down into millions of pieces of micro and nanoplastics.[13, 82, 83] However, microplastics can also come from other primary sources such as textile fibres, pastes, cosmetics, paints, and gels.[84] Animals, birds and fish can eat or choke on these microplastics.[77, 85] Additionally, the ecosystem structure could potentially fail in the long run, due to the sheer amount of non-biodegradable plastic waste in the environment, which can stay there for hundreds of years. Such plastic waste can also accumulate in food chains for human consumption and can be a pathogen carrier.[13, 72, 86, 87]

### *4.3 Contaminated PPE in the environment*

PPEs may become contaminated with microorganisms during patient care or personal use, spread *via* contact, droplets or aerosols.[88] A diverse community of approximately 400 different types of bacteria, mostly toxic, were found in 275 pieces of plastic collected on three beaches in Singapore, and reported to be responsible for coral bleaching, wound infections and gastroenteritis in humans.[89] PPEs in the environment could therefore act as a carrier of COVID-19 or other pathogens to the waste collectors, litter pickers, or public. Under certain conditions, the virus such as SARS Cov-2 can survive up to seven days in the plastic.[90] In many cases, those are persistent pathogens and can survive from a few weeks to several months.[91-94] Indeed, 22 gram-positive bacteria were found on five commonly used hospital products (clothing, towels, scrub suits and lab coats, privacy drapes, and splash aprons), and some of them survived for more than 90 days.[94] In a study,[95] it was found that coronavirus droplets live longer on plastic than other surfaces such as paper or cardboard. It was also showed that textile and PPE play a critical role in bacterial transmission or viral infections.[5, 93, 94, 96-98]

## 5. The Environmental Footprints of Protective Clothing

The textile industry is reported to be the second largest polluter of the environment after the oil industry, and annually half a million tons of microfiber are discharged into the environment.[99, 100] However, the environmental impacts of textiles are unevenly distributed globally due to a dispersed global textile supply chain. The developing countries (mostly in Asia) are hubs of textile manufacturing and bearing most of the burden of these environmental impacts, particularly for natural fibres, such as cotton, wool and silk due to agriculture, farming and processing. In the case of single use PPEs, such environment burden mainly lies on energy and waste, due to its sheer volume production and use.[44, 101] A life cycle analysis (LCA) evaluates the possible environmental impacts of product, processes, and materials, to enable making sound choices for the design, materials or processes involved in manufacturing a product.[102] Within the LCA, a life cycle inventory is considered for quantitative measurement of energy and emissions during the manufacture, use, and disposal. The environmental impacts such as carbon footprints, human toxicity, and eutrophication were quantified based on these inventory outputs.[103] However, the diverse nature of the PPE supply chain makes it difficult to assess the actual environmental impacts.[44]

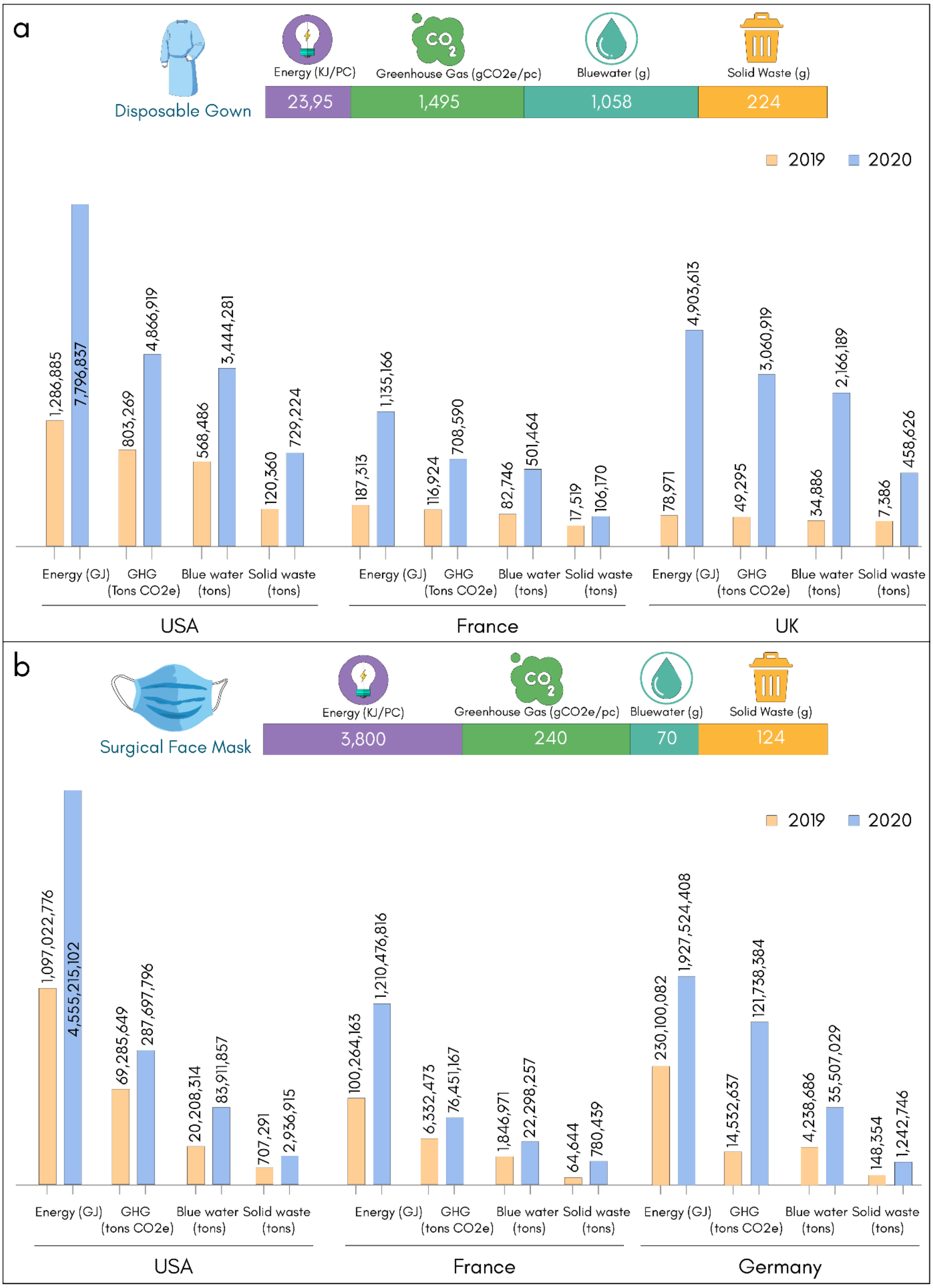

***Figure 6. Environmental Impacts of Personal Protective Clothing based on Six-digit HS Code.*** *a) Disposable gown (HS Code 621010) with weight ~224 g/pc.[104] b) Surgical face mask (HS Code 630790) with weight ~2.45 g/pc.[105] Environmental impacts are calculated and compared based on import data for three major countries in 2019 (Pre-COVID-19) vs 2020 (During COVID-19). Import data is taken in tons from ITC Database.[53, 54] The environmental impact parameters such as energy consumption, greenhouse gas emission, blue water consumption and solid waste has increased linearly to that of importing figures as given in Figure 5.*

In Figure 6a,b, we compare the environmental footprints of PPEs for three main countries in 2019 and 2020. For disposable gown (HS code 621010), the quantity of import has soared for USA (606%), France (6209%) and UK (606%), Figure 6a. Similarly for surgical mask (HS code 630790), the the import quantity increased dramatically for USA (415%), France (1207%) and Germany (838%), Figure 6b. Such dramatic increase in import quantities has resulted in surge for environmental impacts with these products in terms of energy, greenhouse gas emission, water, and solid waste.

### *5.1 Water and Energy Use*

The traditional textile industry is a recognised source of water pollution, and has associated water consumption around 79 billion cubic metres of water in 2015.[106] In general, the water consumed to produce one kg of textile fabrics is between 100 to 150 L/kg, which impacts on the wastewater generated downstream.[107] For example, a study found that between 2012 to 2016, the annual water footprint in the Bangladesh textile industry was found to be ~1.8 billion cubic metres.[108] Additionally, the textiles industry emitted ~1.75 billion $CO_2$e (*carbon dioxide equivalent*) tons globally in 2015,[106] an estimated 8.1%[109] to 10%[110] of total global greenhouse gas emission. In general, the production of nonwoven fabric involves less water consumption and similarly, less water is needed for single-use PPE during their usage. However, it was estimated that two-thirds of $CO_2$e emissions of textile industry is associated with synthetic textiles processing including fibre production, textile manufacturing and apparel production.[44] The high carbon footprint of synthetic fibre production comes from the sources of energy used. For example, China uses coal to produce energy,[111] which will have a ~40% larger carbon footprint than in Turkey and Europe.[112] However, in the life cycle, fibre extraction from fossil fuel has the highest energy use and GHG emission in case of synthetic fibre.[113]

To understand the environmental impact of disposable and reusable gowns, study has been undertaken which includes raw materials to the production of the finished gown and commercial use, cleaning and sterilising of reusable products to the final end-of-life cycle (either incinerated or landfilled as a more prevalent disposal option).[114] Traditionally reusable surgical gowns are made of 100% cotton, followed by cotton-polyester (PET) blends or full PET fabrics[115] – differentiated by woven PET fabric for non-critical zones and knitted PET fabrics in the critical zones[104] with mostly polytetrafluoroethylene liquid-resistant barriers (~70%) or polyurethane breathable barrier membranes (~30%).[116] On the other hand, disposable surgical gowns are made of nonwoven PET and nonwoven polypropylene fabric for the noncritical zones and critical zones, respectively. It was found that the environmental impact of a reusable gown was far less than that of the disposable gown, for example, the use of reusable gowns could reduce natural resource energy consumption (~64%), greenhouse gas emissions (~66%), blue water consumption (~83%), and solid waste generation (~84%).[104] In previous studies between 1993 and 2011, comparative life cycle studies of six reusable and disposable surgical textile were conducted. The result shows that reusable surgical gowns and drapes use more natural resource energy (~200%-300%) and water (~250%-330%), but have lower carbon footprints (~200%-300%) and generated lower volatile organics, and solid wastes (~750%) than disposable gowns and drapes.[114] Additionally, a commercial reusable surgical gown requires ~36.1 g of packaging compared to ~57.8 g for the same for disposable gowns – which eventually translates into a 8% total energy consumption and greenhouse gas emission for reusable surgical gowns compared to 13% for the comparable disposable gown.[104] However, it will be difficult to substitut disposable gowns or any other single-use PPEs of synthetic fibre, unless a recyclable alternative is found, which could meet stringent regulatory requirements for tackling highly infectious diseases like COVID-19.

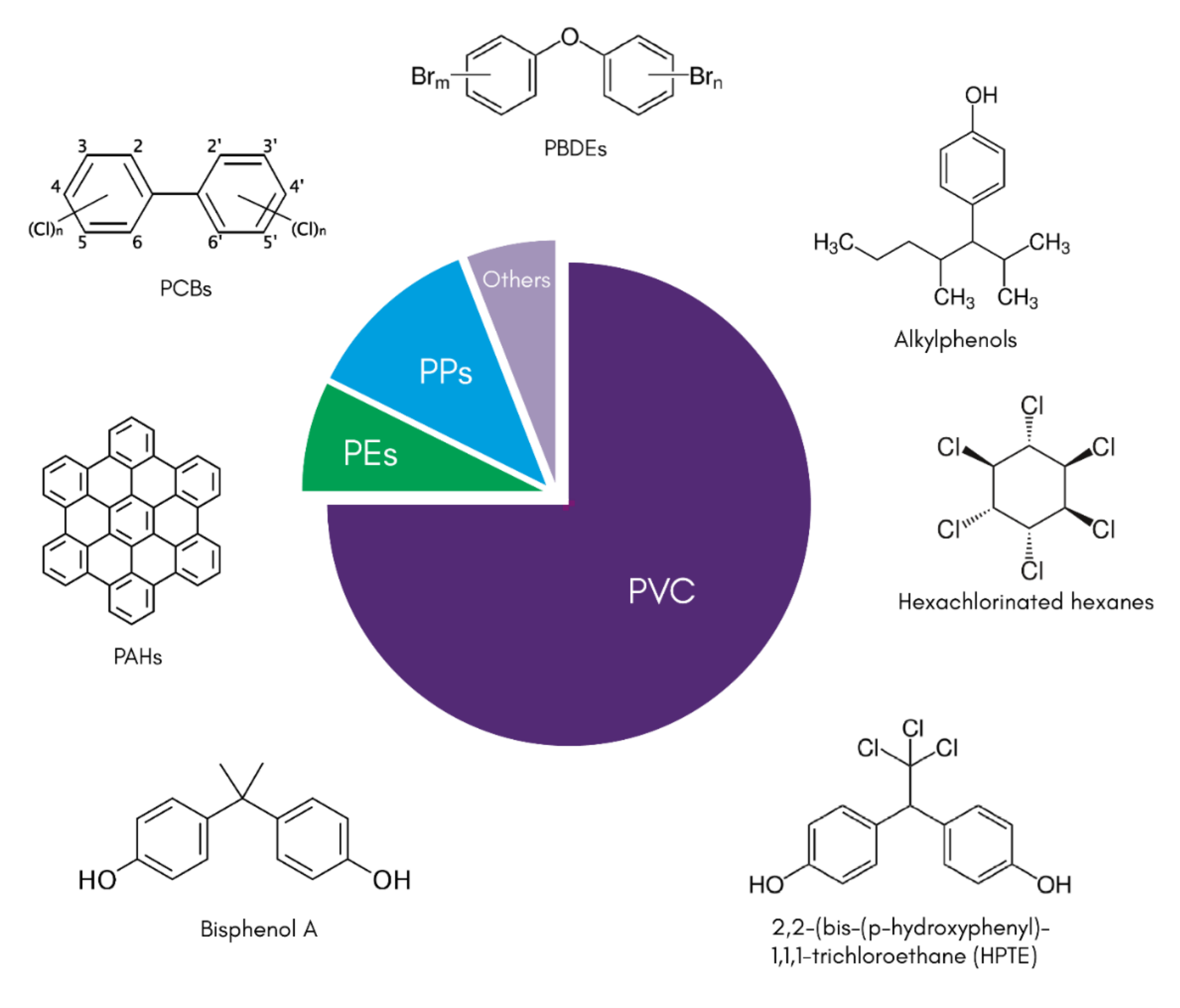

***Figure 7. The use of chemicals additives during PPE manufacturing and end use stages.*** *PPE pollution can contain various additive chemicals, which are usually used to provide certain properties and functionalities to the PPEs. PVC typically requires the most additives (~73% of total production volume), followed by PEs and PPs (10% by volume). Chemical additives are used during manufacturing (fibre spinning, wet processing and finishing) and end use (sterilizing, cleaning and disinfecting) of protective clothing.*

*5.2 Chemical Use*

The use of chemicals for single-use PPEs occurs in the following manufacturing/end-use stages: a) the nature of polymer raw materials and additives, b) chemicals used during processing, c) degradation of polymers in the environment,[117, 118] and d) sterilizing, cleaning and disinfecting.[119] The polymers used in PPEs are usually biochemically inert; however, the polymerisation reaction is, in most cases, incomplete and contains residual monomers, which can be hazardous to human health and the environment.[120] The fraction of the residual monomer varied from ~0.0001% (100 ppm) to ~4% (40,000 ppm), and depends on the type of polymer, polymerisation technique and other variables.[121] With its diverse polymer types, PPE pollution can contain various additive chemicals, which are usually used to provide certain properties and functionalities to the PPEs.[122, 123] More than several thousand different additives exists for plastic polymers, but these are unevenly distributed. PVC typically requires the most additives (~73% of total production volume), followed by PEs and PPs (10% by volume), Figure 7.[124] These additives are organic chemical compounds like polychlorinated biphenyls (PCBs), polycyclic aromatic hydrocarbons (PAHs), persistent organic pollutants (POPs), organochlorine pesticides (2,2′-bis(p-chlorophenyl)-1,1,1-trichloroethane, hexa-chlorinated hexanes), polybrominated diphenyl ethers, alkylphenols and bisphenol A, other additives or plasticisers and associated degradation products in the range of concentration from sub ng $g^{-1}$ to µg $g^{-1}$.[81, 118, 125] These are persistent toxic chemicals in the marine environment, which can leach out and adhere to the surface and add further contamination.[126] The release of these degradation products could occur during production, use and in the end of life phase.[127] When plastic materials are exposed to the dissolved chemicals already present in the ocean environment[118], it can also release harmful chemicals as evident in the nutrient-rich stomach oil of seabirds over time,[47, 117, 128] which may negatively affect reproduction through disrupting hormone release and may have long term genetic effects in birds[117] and other marine animals.[129, 130] The transfer of these chemicals from plastic materials in a living organism could be by ingestion, excretion, as a direct source, dietary or dermal transfer.[81] The debate of the use of bioplastics (e.g PLA),[131, 132] as a substitution of petrochemical-based plastics, is also significant, as the sources are mainly sugar and starch materials – a direct competition to food crops, and also include chemicals and additive during manufacture.[133]

The traditional textile industry is reported to use more than 8,000 chemicals in its many and varied manufacturing processes, and the persistence of the materials in the environment is the ongoing challenge.[134] Similarly in the manufacture of PPEs, chemicals are used in the spinning of fibre (solvents, lubricants), processing (chlorine for bleaching, dyes in dope dyeing, flame retardant, water repellents, antibacterial finish etc.), fabric production (epoxy or other resins).[44] However, the actual amount of chemicals required to produce a kg or a piece of protective clothing is unknown. The sterilizing, cleaning and disinfecting of PPEs also uses chemicals such as hydrogen peroxide as a disinfectant.

Ethylene oxide for sterilisation is also recommended for the use of recyclable PPEs. A list of products that can be used is also specified by USEPA, particularly for COVID-19.[135] The use of anti-microbial finishing in protective clothing is discussed elsewhere.[2]

In general, it appears that the chemical footprint of single-use nonwoven protective clothing is comparatively lower than the traditional clothing. However, there are still many unknown factors, such as the production environment, pollution mitigating technology, and waste treatment facility. These all could lead to higher environmental impacts, and health and safety risk to the workers, producers and users. Although the physical and chemical toxicity of microplastics due to contamination, consumption and other factors on human are yet to be fully determined,[70, 128] nevertheless it has been reported that depending on the pre-existing health conditions, microparticles from plastic can cause alterations in chromosomes which may lead to infertility, obesity, and cancer.[71, 101]

### *5.3 Waste generation*

The UN Economic Commission for Europe (UNECE) identified the textiles industry as a significant contributor to plastic entering into the ocean.[136] Plastics represent ~13.2% of total municipal solid waste generation in 2017 in the US out of 35.4 million tons of total waste. The American Chemistry Council analysed the presence of plastics in municipal solid waste from 1960 to 2017[90] and found that ~13.2% (~35.4 million tons) of the total waste generated in the US was plastics, mostly polyethylene and polypropylene.[90] In these six decades, 0 to 9% of the municipal plastic waste was recycled, ~2% to ~17% were recovered for energy and ~75% to ~100% was landfilled in the 10 year period .[90] In addition, the total plastic waste in the waterbodies arising from land-based waste, particularly in densely populated or urban areas such as Tokyo, Nagoya and Osaka was high.[81] The problems associated with these microplastics are increasingly pervasive and are found in seafood, beer, honey, table salt and bottled mineral water.[69] After domestic or hospital use, single-use PPEs are discarded either into landfill and may impact on landfill seepage in future years.[2]

It is estimated that without systematic change, 12 million tons of plastic litter will end up in the environment such as landfill and ocean.[18] and will contribute greenhouse gas emissions up to ~5% of the global carbon budget by 2050.[137] The effect of this plastic accumulation in nature could be multifold. If land pollution is considered then the blockage of the sewage system can increase the risk of flooding,[138] can be a breeding ground for vector-borne zoonotic diseases (e.g. *Aedes* sp. mosquitoes, as a vector of dengue and zika),[139] and can degrade soil and be responsible for poor crop development.[140] Additionally, plastic debris can reach the aquatic ecosystem through various water channels such as a sewage line, wastewater treatment plant, rivers and ocean and can reach the furthest areas of the Earth such as Antarctica.[101]

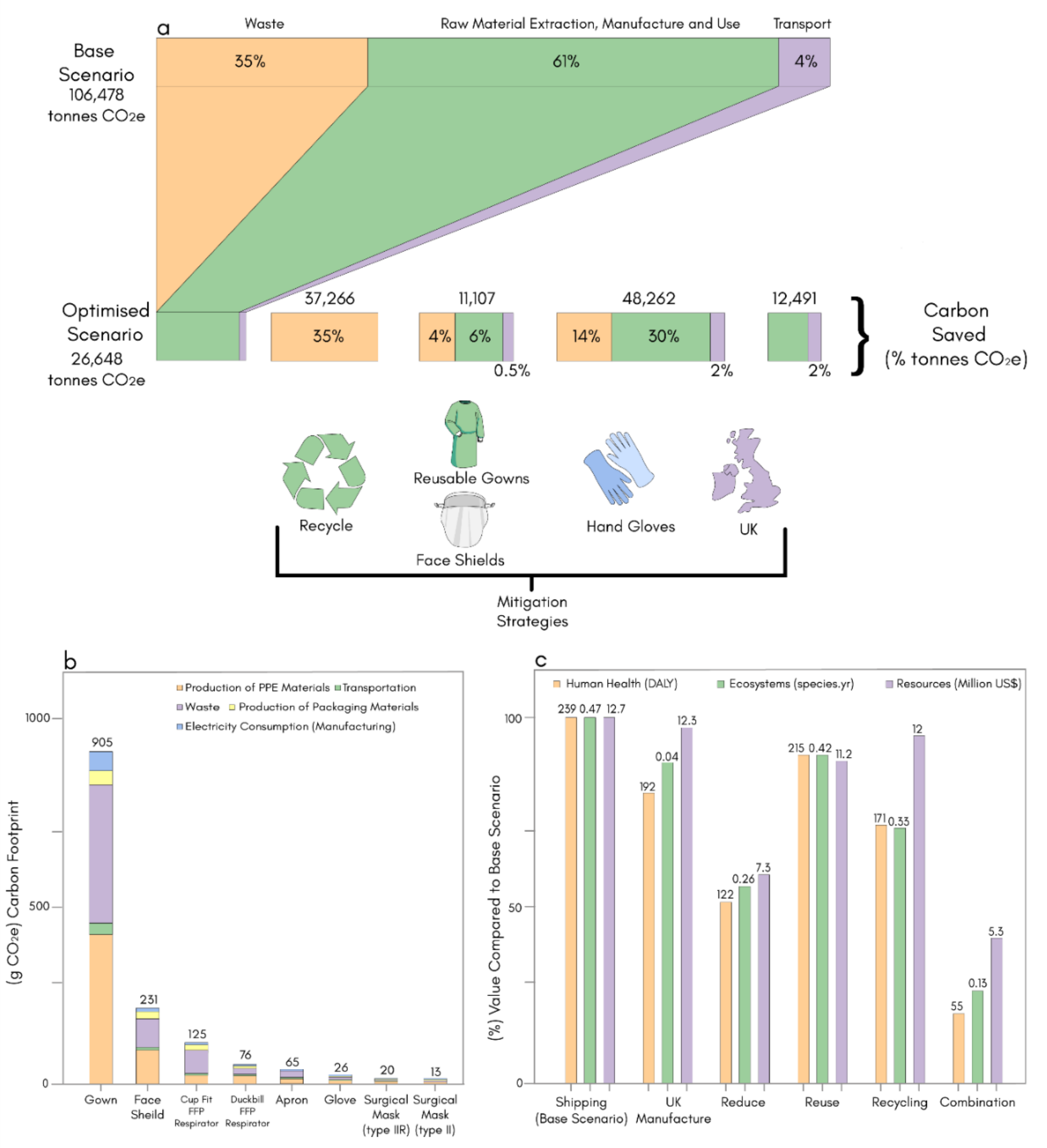

***Figure 8. Mitigation of environmental impacts of personal protective equipment (PPE). a) Carbon footprints** of PPEs used by the NHS in UK from February to August 2020 of the COVID-19 pandemic in the base and optimised scenarios (UK manufacture, eliminating glove use, reuse of gowns and face shields, recycling). b) Carbon footprint of individual single-use PPE items with process breakdowns (production of PPE materials, transportation, waste, production of packaging materials and electricity consumption during manufacturing). c) Environmental impacts (endpoint categories) of alternative scenarios for PPEs used by the NHS in UK from February to August 2020 of the COVID-19 pandemic. The base scenario includes shipping, single-use and clinical waste. Alternative scenarios are the use of UK manufacturing, reduce (zero glove use), reuse (reusable gown, reuse of face shield, all other items single-use), recycling and combination of measures. (DALYs= disability-adjusted life years, loss of local species per year in species.year, and extra costs involved for future mineral and fossil resource extraction in US $).(reprodcued with permission from[22] )*

## 6. Future Directions

### 6.1 Reduce, Reuse, Recycle and Local manufacturing of PPEs

The use of single-use PPEs will not be a sustainable practice into the future.[1] Reuse of PPEs is an option, and are already used in many settings, for examples face shields and reusable gowns in operating theatres. Reusable face shields and gowns were found to lower environmental impacts up to five-fold compared to a single-use version. [59, 141] The UK and Wales government has reiterated not to use single-use PPEs wherever possible to manage their environmental impact and to support recycled and reusable alternatives.[63] A detailed analysis of these approach will be required so that reusable PPEs do not compromise the primary function of protecting health. PPE sterilisation on a large scale will be needed for reuse, which is possible through hydrogen peroxide vapour, ultra-violet or gamma-radiation or through other spray-on disinfectants.[59, 142] In a recent study[22] on the most commonly used PPE items by the National Health Services (NHS) in UK, the overall environmental impacts of masks, gloves, aprons, gowns, and face/eye protection were evaluated. From February to August 2020 of the COVID-19 pandemic, the total carbon footprint of all PPEs supplied was 106, 478 tonnes $CO_2$e for base scenario, in which ~61% was derived from raw materials extraction, manufacture and use, 35% from waste and 4% from the transportation, Figure 8a, b. However, carbon foot prints could be reduced by 11%, 46%, 10.5% and 35% *via* UK manufacturing, reduce PPE use (eliminating gloves), reusing and recycling of PPS PPE, respectively (Figure 8a, c).[22] In addition, PPEs will be in high demand into the foreseeable future and the investment in new PPE materials at a global level is key. A multi-disciplinary team with technical expertise including material science, biomedical science, environmental science and product engineering is essential to tackle the PPE pollution problem.

### 6.2 Removing Supply Chain "Bottlenecks"

Although initiatives are emerging encouraging local production, particularly for emergency supplies, it is still a challenge due to the fragmented nature of the supply chain and the need for rigorous quality assurance. The process of sourcing materials, designing assembly processes, machining and scaling up the production, quality and testing procedures, certifications, etc. will be required in all cases.[48] In addition, transport and shipping, containers, limited workforce all will be significant factors in managing the complex global supply chain of textiles.[43]

### 6.3 Waste Management

The pandemic has stressed the solid waste management infrastructure globally, highlighting the supply chain difficulties across PPE manufacture, demand-supply, use, logistics and disposal.[1] Even in "normal times" the efficient management of waste is a significant challenge;[44, 75, 143, 144] and in most developing country there are fewer management options with main choices being landfill or open burnings.[145]

Due to the highly contagiousness nature of the COVID-19 virus, many countries classified all hospital and domestic waste as infectious,[145] which should be incinerated at high temperature followed by landfilling of the residual ash.[146] Some larger economies were able to manage this option; for example, China deployed mobile incineration facilities around Wuhan to tackle infectious waste.[147] But in most cases, the significant increased consumption of single-used PPEs along with other medical waste due to the pandemic will most likely overload waste management.[14] In general, the basic principles of waste management strategy are: reduce-reuse-recycle and these fundamentals should be applied to PPEs. Also, within the circular economy philosophy these principles should guide policy development during and after the current pandemic. National policy should encourage recycling, incentivise adoption and embed "cheap" product pricing. The economic model will promote the adoption of green chemistry and technology, safe process, life cycle analysis.[1] In addition, strategic policy options can be implanted based on the share of use of PPEs or based on the individual carbon footprint. For examples, gloves are responsible for 47% of the PPE carbon footprint, and their usage is a key area for innovation and could be prioritised.[59] In reality, sustainable management of PPE waste will be a crucial challenge[1, 63, 75, 148] towards achieving the United Nation's Sustainable Development Goals (SDGs) such as SDG 3 - good health and wellbeing, SDG 6 - clean water and sanitation, SDG 8 - decent work and economic growth, SDG 12 - responsible consumption and production and SDG 13 - climate action.[57] Traceability of production of PPEs and corresponding waste management perhaps could be a key for unlocking these challenges.

### 6.4 Smart and Sustainable Materials

As discussed before, nonwoven PE and PP fabrics are the main raw materials for single use PPEs, based on various spunbond-melt-spun materials. Such materials would be very difficult to replace, particularly for hygiene and health requirements. However, it is possible to use in combination with some natural, regenerated or biodegradable fibres,[131, 132, 149] which can then be either biodegradable and/or could provide reusable properties. In addition, the substitution of some chemicals/additives currently used in the production of PPEs provides an opportunity for an integrated approach to eliminating persistent and damaging materials. Additionally the use of new materials such as graphene[150] for manufacturing PPE could potentially help moving towards sustainable products with enhanced mechanical properties. However, substitution of these chemicals used in the production of PPEs should be enforced by legislation and regular monitoring. Coupled to these local changes, wider scale import restrictions could also help to accelerate the acceptance of a greener philosophy in selecting raw materials and chemicals of PPEs.

Smart PPE, has also gained significant attention in recent years due to their ability to improve workplace safety and achieve operational excellence. Such PPEs are usually connected to wearable devices, and continuously track movement and monitor vital physiological conditions including temperature, heart rate and breathing rate. Smart PPE can capture and track thousands of different data points, which can be used to address any number of safety concerns, everything from fever to heat exhaustion to fatigue to improper lifting motions. Smart wearable e-textile technologies[151] could be integrated with protective clothing to produce truly "Smart" wearable medical clothing. In previous studies,[152-155] we reported washable, durable, and flexible graphene-based wearable e-textiles, which are highly scalable, cost-effective, and potentially more environmentally friendly than existing metals-based technologies. It could potentially lead to manufacturing of smart, sustainable and reusable personal protective clothing with less environmental impacts.

**Acknowledgement**

The authors respectfully acknowledge all the front-line workers ("Heroes") around the world for the wonderful care and support they are providing daily during the current Covid-19 pandemic. This research was supported by E3 Research England Funding (UK).

**Authors Biography**

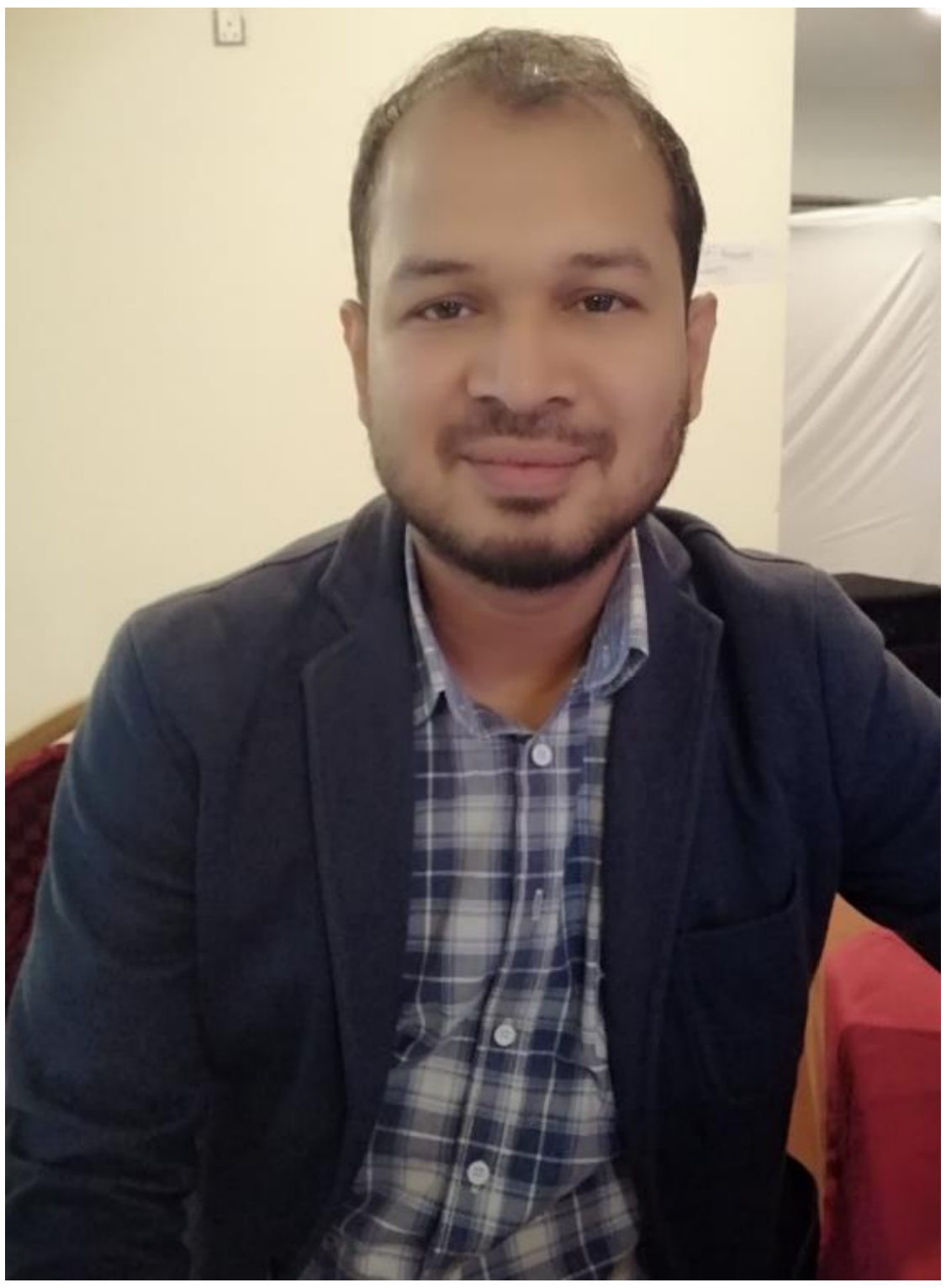

**Dr Mohammad Abbas Uddin** is an Assistant Professor at Bangladesh University of Textiles. He is also working on future skills development of textile graduates in collaboration with a2i, and as Assistant Director in skill development project funded by ADB. Dr Abbas is one of the authors for producing 'National Chemical Management Guideline for Textile industry' 2021. He has over 18 years of experience, specialising in Textile wet processing, value chain and environmental sustainability. He holds a PhD from the University of Manchester, MBA from IBA, University of Dhaka, and Masters from Curtin University. He is a Chartered fellow of Textile Institute, UK.

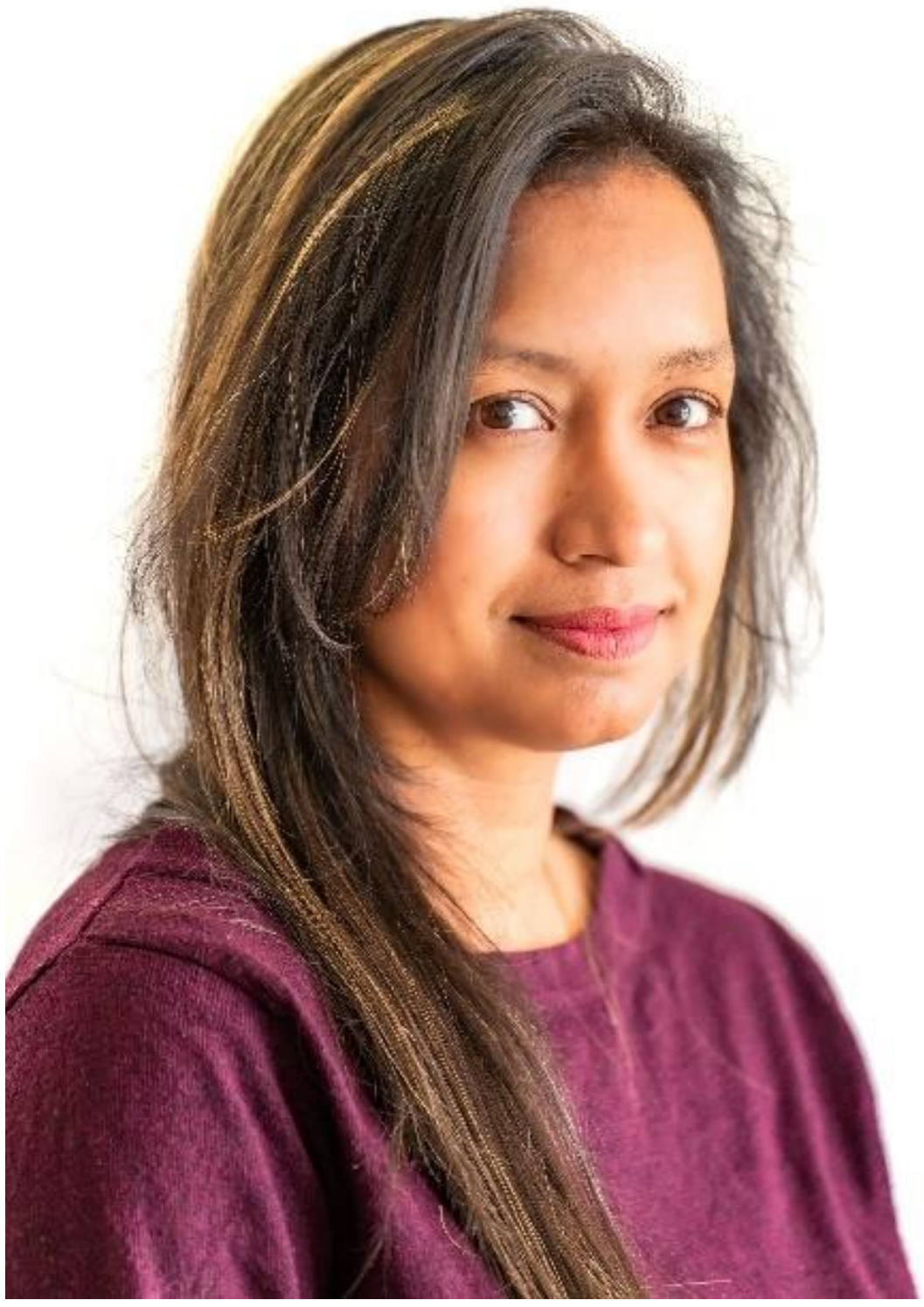

**Dr Shaila Afroj** is Senior Research Fellow at the Centre for Print Research, UWE Bristol, where she investigates graphene and other 2D materials-based technologies aimed at developing next generation wearable electronics textiles and sustainable functional clothing. Prior to that, she worked as a Research Associate at National Graphene Institute, the University of Manchester after completing her PhD from the same university. She has about 13 years of industry (including multi-nationals companies like C&A and Intertek) and academic experiences realated to advanced materials, wearable electronics and fashion textiles.

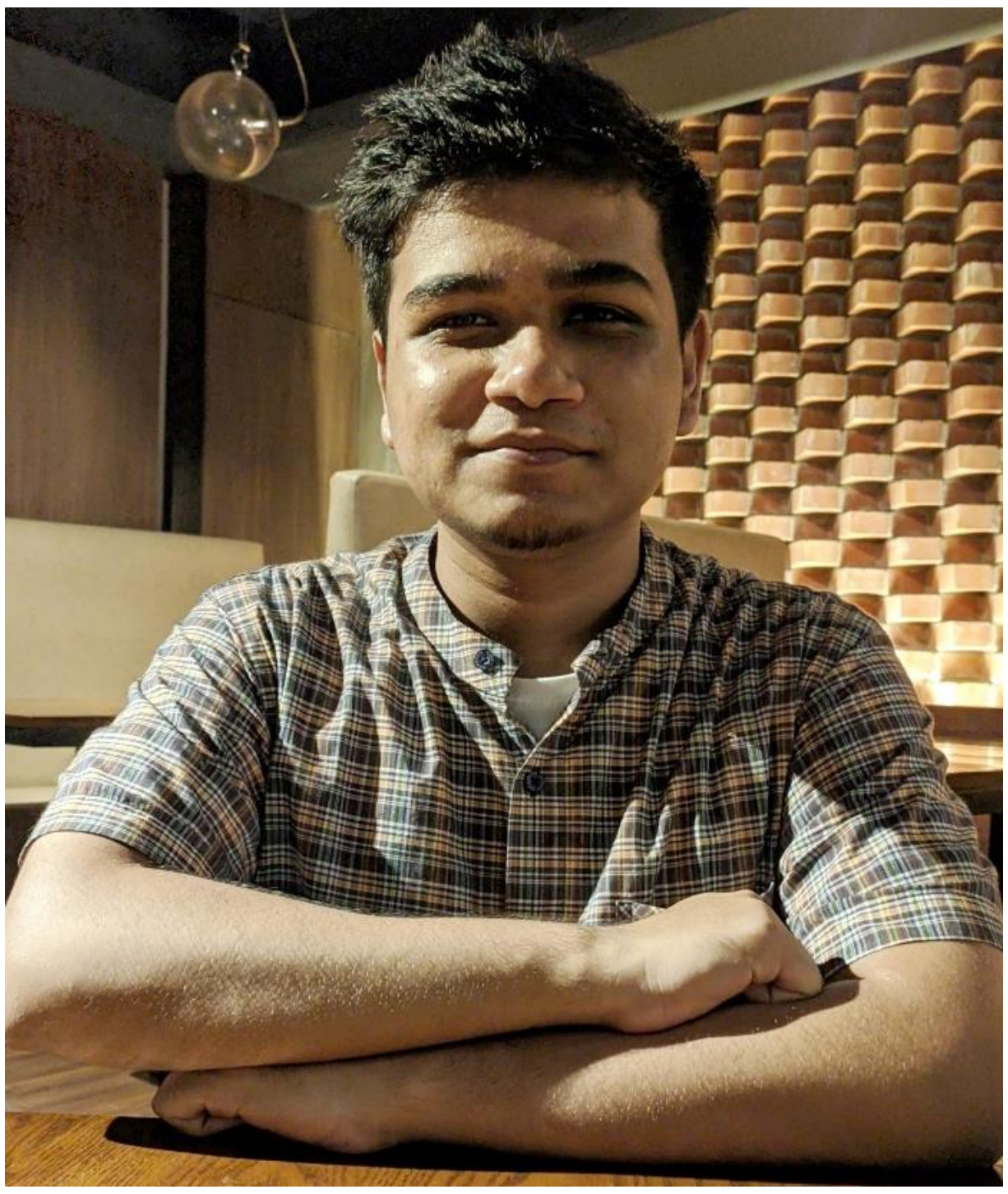

**Tahmid Hasan** is BSc Textile Engineering student at the Department of Environment Science and Engineering of Bangladesh University of Textiles (BUTex). He has research interests into environmentally sustainable clothing and wearable electronic textiles. He is currently part of a BUTex research team working on graphene-based wearable e-textiles.

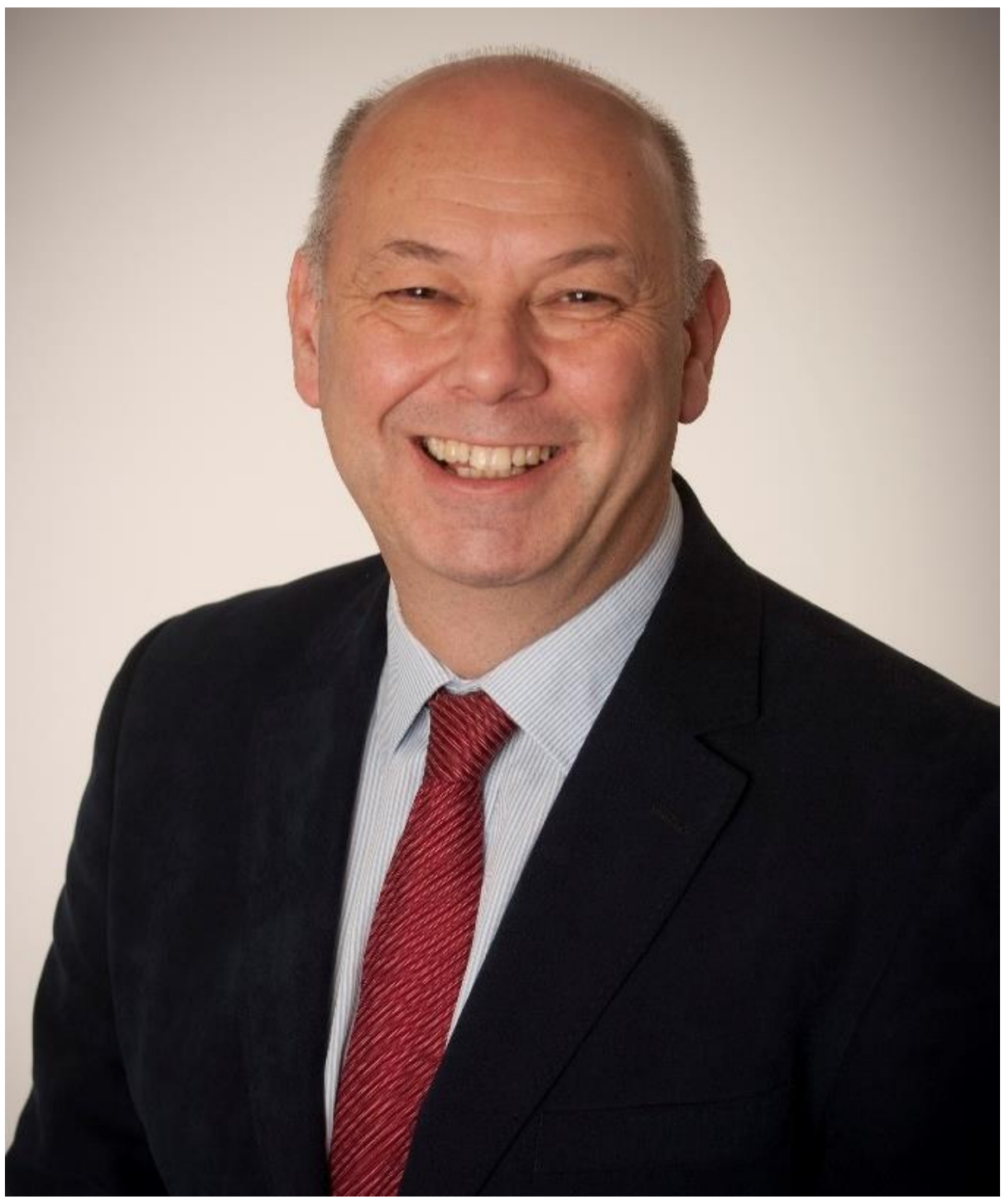

**Professor Chris Carr** is Professor of Textile Technology and Director of the 3D Weaving Innovation Centre at the School of Design, University of Leeds. His main interests are in the modification of fibrous materials by dry and wet chemical and biotechnological modification to improve product performance. Research has focused on easy care finishing of cotton and wool, laundering processes, conservation science, technical textiles, hair processing, surface chemistry, novel colouration and healthcare textiles/materials. He has published widely, is a member of several Editorial Boards for international research journals and is a Liveryman in the Dyers Company in London.

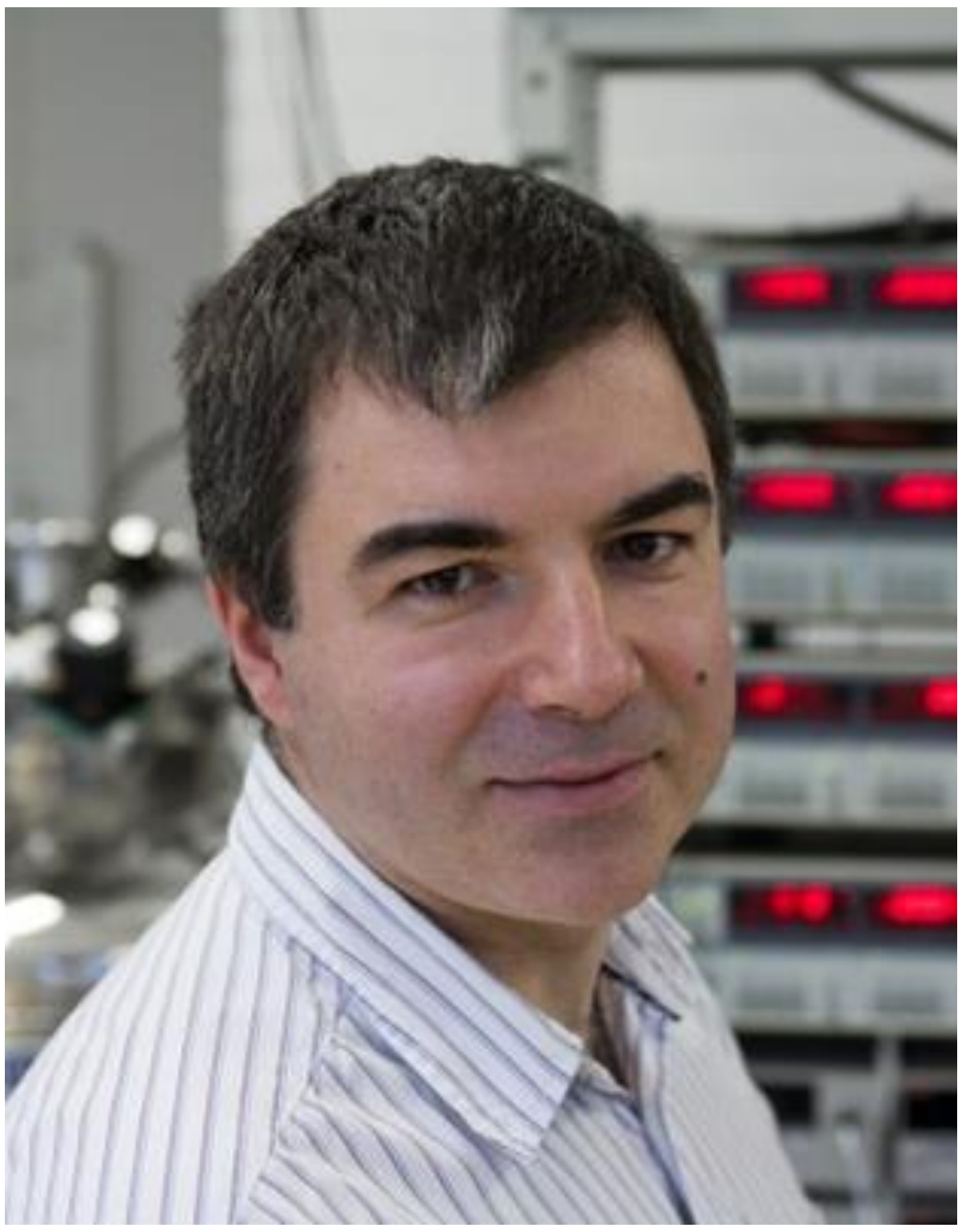

**Professor Sir Kostya S. Novoselov** is an condensed matter physicist, specialising in the area of mesoscopic physics and nanotechnology. He is currently Tan Chin Tuan Centennial Professor at National University of Singapore with broad research interests from mesoscopic transport, ferromagnetism and superconductivity to electronic and optical properties of graphene and two-dimensional materials. He also has got a vast background in nanofabrication and nanotechnology.

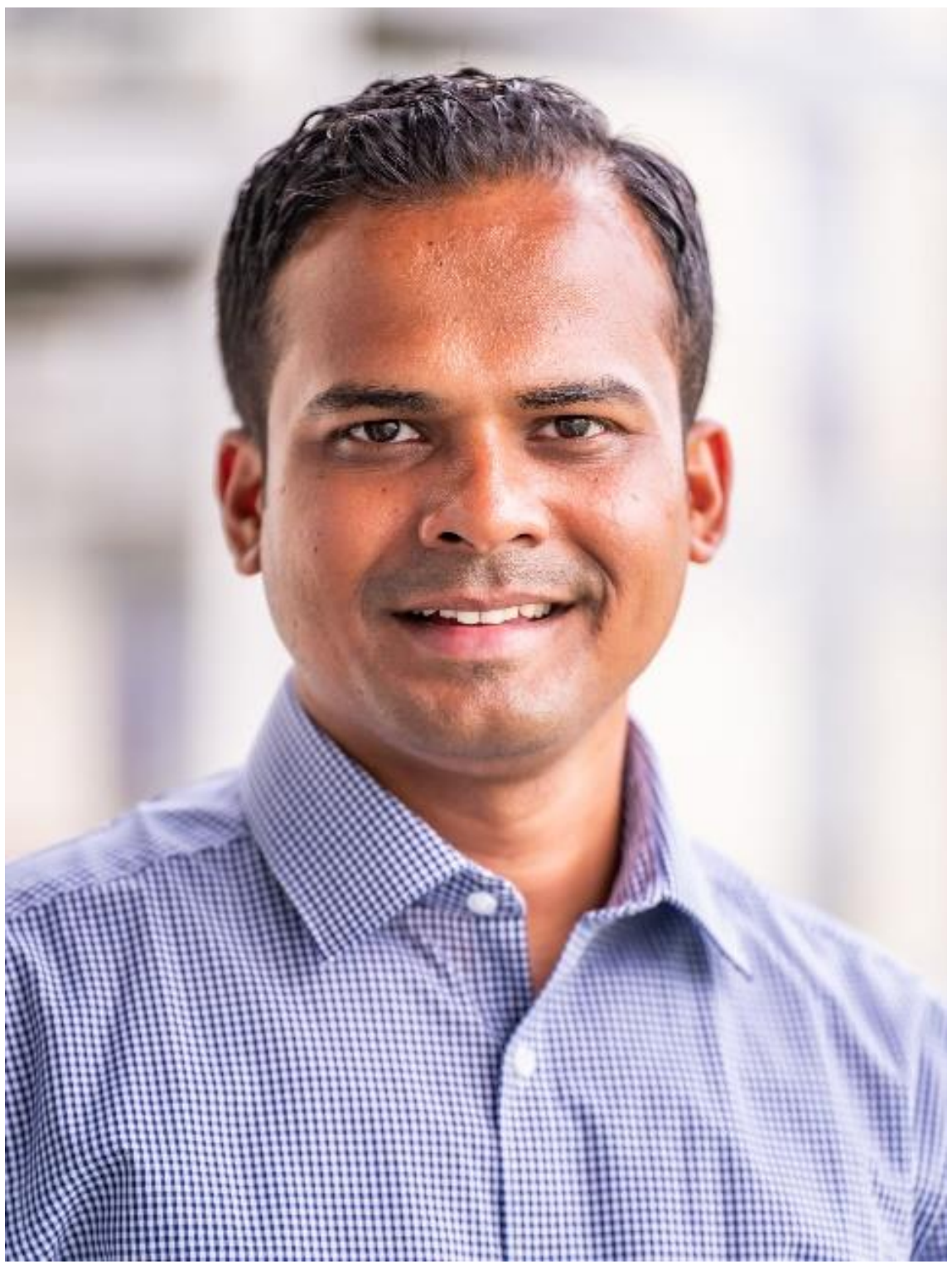

**Dr Nazmul Karim** is Associate Professor at the Centre for Print Research, UWE Bristol. He is currently leading a research team to investigate into graphene and other 2D materials-based technologies for developing next generation wearable electronic textiles, environmentally sustainable functional clothing and fibre-reinforced composites. Prior to that, Dr Karim was a Knowledge Exchange Fellow (graphene) at the National Graphene Institute of University of Manchester. He has about 13 years of industry and academic experiences in graphene and textile related technologies, and a passion for getting research out of the lab and into real world applications.